\documentstyle[preprint,aps,eqsecnum,epsf,tighten]{revtex}
\begin{document}
\preprint{\baselineskip 18pt{\vbox{\hbox{SU-4240-662} \hbox{OUTP-98-12P}}}}
\title{Domain Formation in Finite-Time Quenches}
\vspace{15mm}
\author{Mark Bowick}
\address{Department of Physics, Syracuse University,\\
Syracuse, NY 13244-1130, U.S.A.}
\author{Arshad Momen}
\address{Department of Physics,
Theoretical Physics,\\ University of Oxford, 
1 Keble Rd., OX1 3NP, U.K.}
\vspace{15mm}
\maketitle
\begin{abstract}
We study the formation of domains in a  continuous phase transition
with a finite-temperature quench. The model treated is the $\Phi^4$
theory in two spatial dimensions with global $O(2)$ symmetry. 
We investigate this using
real-time thermal field theory, following Boyanovsky and collaborators, 
and find that 
domain sizes appear to be smaller
than those produced in an instantaneous quench in the tree-level
approximation.  We also propose that a more physical picture emerges 
 by examining the two-point functions
which do not involve any cutoff on the short wavelength
Goldstone  modes. 
\end{abstract}
\vspace{5mm}

\newcommand{\be}{\begin{equation}}
\newcommand{\ee}{\end{equation}}
\newcommand{\bea}{\begin{eqnarray}}
\newcommand{\eea}{\end{eqnarray}} 
\newcommand{\real}{{\rm l}\! {\rm R}}
\newcommand{\ra}{\rightarrow}
\newcommand{\tr}{{\rm Tr}\;}
\newcommand{\pt}{\partial}
\newcommand{\bt}{\beta}
\newcommand{\half}{\frac{1}{2}}
\newcommand{\cu}{{\cal U}}
\newcommand{\om}{\omega}
\newcommand{\tq}{\tau_Q}
\newcommand{\thr}{\frac{1}{3}}
\newcommand{\lam}{\lambda}

\section{Introduction}

The formation of domains in phase transitions, and the
subsequent (growth) dynamics of these domains is an important
topic arising in a variety of contexts
ranging from early universe cosmology to condensed-matter physics.
As a result, domain formation  has been much investigated 
from both the classical~\cite{bray} and 
quantum mechanical~\cite{rr} points of view. 

There are two principal mechanisms of domain formation, namely 
a) Bubble-Nucleation and b) Spinodal Decomposition. 
Strongly first order phase transitions typically proceed via bubble-nucleation  
while spinodal decomposition  is often observed 
in second order phase transitions. For a weakly first 
order phase transition, however, the idea that the 
phase transition is driven by the formation of bubbles is not so clear.
In fact, this type of phase transition
can be driven by spinodal decomposition as well. Nevertheless,
domain-type structures appearin both mechanisms. 

The growth process of such domains at late times (also known as phase
ordering in the literature) 
is relatively well-understood \cite{bray} while the study of
the formation process of the defects associated with these domains has
begun only recently\cite{zurek}. The initial size of the domains
formed in continuous phase transitions is of physical interest. 
For phase transitions that generate topological  defects this domain
size also determines the initial defect density. 

The earliest attempt to determine the initial domain size, in the cosmological
context, is due to Kibble \cite{kibble,HiKi95}.  Along with a {\it precise}
formulation of the mechanism by which topological defects are produced
in phase transitions Kibble made a rough estimate, using a
thermodynamic equilibrium picture,  of typical
domain sizes. This was done  by determining how big an 
ordered domain had to be in
order to remain stable under thermal fluctuations
to the disordered phase.  But domain
formation is inherently a non-equilibrium process and, therefore, one must be
careful about such naive arguments.  
A more refined viewpoint, based partly on  non-equilibrium ideas, 
has been proposed by Zurek \cite{zur}. 
In the latter picture one focuses on the competition between the
applied quench rate and the intrinsic relaxation rate of the order 
parameter. Defects will freeze out when the order parameter cannot adjust 
rapidly enough to follow the quench. The correlation length at this
freeze-out time determines the characteristic size of
ordered domains and hence the typical separation between the resultant
topological defects. The non-equilibrium ingredients of this picture
imply that the domain size may differ differ significantly from the
simple predictions following from thermal stability. Zurek
and co-workers have also  given  numerical evidence for this picture using
computer simulations \cite{laguna}.

The processes of domain formation and growth in quantum field
theory at finite temperatures have been investigated by Boyanovsky and
collaborators  \cite{boy} using the real-time
formalism\cite{LeBelac,CaHu87}. This formalism is particularly suitable for
time-dependent processes like domain formation for which   
the more popular (imaginary time) Matsubara frequency method is 
rather cumbersome to use. Applying these techniques for scalar
field theories with quartic interactions Boyanovsky et al. \cite{boy}
managed to reproduce the well-known Cahn-Allen growth laws in the
classical theory of phase-ordering\cite{bray}. The phase
transition in such systems was induced by adopting  a 
sudden quench of the heat bath. Though the quench is performed on the 
heat-bath rather than on the system, Boyanovsky et al. assumed that the resulting
effect was mocked up by taking the mass-function for the scalar
fields to be a step-function with a change of sign. 
Consequently, unstable long-wavelength modes appear in the
theory which grow  exponentially with time and lead to the 
appearance of domains.

Scalar field theories admitting global $O(d)$ symmetry 
admit topological defects in various dimensional spaces. 
These global topological defects contain singularities of the order parameter which are the
zeros of the $O(d)$ symmetric scalar field. The density of such defects
is then related to the density of the associated zeros and these can
be determined, in a Gaussian approximation, by the Halperin-Mazenko-Liu 
\cite{halperin,mazenko} formula, namely 
\be
n(t) \sim \left|\frac{W''(0,t)}{W(0,t)}\right|^{d/2},
\label{0.1}
\ee
where $W(x,t)$ is given by the diagonal equal-time two point-function $ \langle
\Phi_a(x,t) \Phi_b(0,t) \rangle \equiv \delta_{ab} W(x,t)$ and the double
primes denote the derivative with respect to the radial variable $ r \equiv
|r|$.  Once the two-point function is determined one can use
(\ref{0.1}) to find the defect density \cite{gill}. Though one
might question the validity of the Gaussian approximation in an
interacting system, for $O(n)$ symmetric field 
theories it can be viewed as a first order result in a more general scheme \cite{wickham}.

To be precise, the use of a time-dependent 
mass-function is flawed as the temperature dependence
of the masses is obtained using an equilibrium picture which is
definitely not valid with a sudden quench. 
It is thus interesting to adopt a quench  ``slow'' enough that 
the approximation to equilibrium remains  valid. 
Though one can expect the late time behavior of the system to be
independent of the details of the quench process, this will not be
true of the early time
behavior of the system. Our goal is to follow the
system subsequent to the change in the mass function which we assume
to be linear. Such a linear quench has recently been investigated
in \cite{karra}, where it is also found that the defect densities are
affected by the existence of a finite time quench - indeed
being smaller. Though this result seems correct on physical grounds -
its validity can be questioned due to the
approximations used. On the other hand,
the zeroth order equation of motion can be solved
exactly and  one can find 
the two-point function for the unstable modes, following Boyanovsky
et. al. \cite{boy}. 
We find that the domain sizes are larger compared to those appearing
in a sudden quench,
as found by  \cite{karra} but only during the early part of the
quench  and towards the end of the quench and thereafter 
the domains grow more slower in the finite-time quench. 
This result seems highly counterintuitive as a very
slow quench will barely produce defects.  On the other hand, the results of
\cite{karra} show a larger domain size with increasing quench time and
hence a lower density of defects.

We argue that this puzzle originates from an inappropriate identification of
the Goldstone modes which are associated with the defects - they are
not affected by the quench directly. Indeed, we
show that by adopting a polar parametrization of the global $O(2)$ model
discussed in \cite{karra} - the emergent  picture is consistent with
the scenario proposed by Zurek.

The outline of the paper is as follows. In section II we briefly  review
the results of Boyanovsky et.\,al \cite{boy},
for the case of a sudden quench, to highlight the various
approximations used. In section III we generalize to the case of 
 a finite-time quench in their scenario showing that to zeroth order
the domain sizes are smaller. In section IV we suggest that
more physical results are obtained by studying 
the correlation functions for the polar fields.

\section{Predictions for a sudden quench}

Let us start from the well-known $SO(2)$ symmetric $\phi^4$ theory in 2+1 
dimensions, which admits global vortices since $\pi_1 ( S^1)= Z$. 
Most of the results below can be found in the paper of Boyanovsky
et.\,al.  \cite{boy}.  
The Lagrangian for the system is given by
\be
L =\int d^3x [ \half \pt_\mu \bbox{\phi}\cdot \pt^\mu \bbox{\phi} - 
\half m^2(t) \bbox{\phi}\cdot \bbox{\phi}
- \frac{1}{4 } \lambda (\bbox{\phi}\cdot \bbox{\phi})^2 ]
\label{C1}
\ee
where $\bbox{\phi} \equiv \left(\begin{array}{l} \phi_1 \\
\phi_2 \end{array} \right)$ is a two component vector 
under $O(2)$ and we have a time-dependent mass function $m(t)$, given by 
\be
m^2(t) = m^2_i \Theta(-t) - m^2_f \Theta(t).
\label{C2}
\ee 
Due to the $O(2)$ symmetry, the two-point function for this model is
diagonal $\langle \phi_a \phi_b \rangle  = \delta_{a\,b} \langle \phi
\phi\rangle $, where  $\langle \phi \phi \rangle$ is the two-point
function for a single scalar field. Thus, the results for the $O(2)$
will coincide with those for a single scalar field, treated by
Boyanovsky et. al. \cite{boy}.

The zeroth order equation of motion yields 
\be
[\frac{d^2}{dt^2} + {\bf k}^2 + m^2(t) ] {\cu}_k(t) = 0,
\label{C3}
\ee
where ${\cu}_k(t)$ is the single particle wavefunction, as defined in
the  appendix.

With the mass-function defined by (\ref{C2}), at times $ t < 0 $ , 
Eqn. (\ref{C3}) reads, 
\be
[\frac{d^2}{dt^2} + {\bf k}^2 + m^2_i ] {\cu}_k(t) = 0,
\label{C4}
\ee
so that one has as a solution of the form
\be
\cu_k (t) = e^{-i \om(k) t},
\label{C5}
\ee
with $\om^2(k) = {\bf k}^2 + m^2_i $, which we can treat as an
``initial condition''.

In the sudden quench scenario, $m^2(t)$ changes instantly and at 
times $ t> 0$, Eqn. (\ref{C3}) reads
\be
[\frac{d^2}{dt^2} + {\bf k}^2 - m^2_f ]\, {\cu}_k(t) = 0. 
\label{C6}
\ee
Clearly, the modes with 
${\bf k}^2 < m^2_f$ are unstable and they are responsible for the 
formation of domains \cite{weinberg}. For these modes, 
the solutions to Eqn.(\ref{C6}) are given by
\be
\cu_k(t) = A_k e^{W(k) t } + B_k e^{-W(k)t},
\label{C7}
\ee
with 
\be
W(k) = \sqrt{ m^2_f - {\bf k}^2}.
\label{C7.5}
\ee

As we are dealing with a second order differential equation,
the solutions (\ref{C5}) and (\ref{C7}) and their first derivatives
are required to match  
at $t=0$ , leading to 
\bea
A_k &=& \half ( 1 - i\frac{\om(k)}{W(k)}) \nonumber \\
B_k &=& ( A_k)^*, 
\label{C8}
\eea
where $*$ denotes complex conjugation. For brevity, hereafter  
we will suppress the 
functional dependence of $\om $ and $W$ on 
$k$.  

Given $\cu_k(t)$, one readily finds for $t>0$ , using (\ref{B5}), 
\be
G(\bbox{r},t) =
\frac{1}{2} \int \frac{d^2k}{(2\pi)^2 2\om} \coth(\half \bt_i \om) 
\left[\left( 1+ (\frac{\om}{W})^2 \right) \cosh 2W t + \left(
1-(\frac{\om}{W})^2\right) \right]
e^{i\bbox{k.r}}.
\label{C9}
\ee
Note that in the above integral only the unstable modes are taken into 
account, following \cite{boy}.

The above integral, however, includes the contributions that are already 
present before the quench $i.e.$ for $ t \leq 0$ and hence is not 
appropriate for
studying the {\it growth} of domains. One therefore must
subtract the contribution that is present at the beginning 
of the  quench. For the unstable mode with wavenumber $k$ , the 
growth is given by the function 
\be
\tilde{S}_k (t) \equiv  |\cu_k(t)|^2 - |\cu_k(0)|^2.
\label{C10}
\ee
In coordinate space this translates into 
\bea
\tilde{G} (\bbox{r},t) &=&\int \frac{d^2k}{(2\pi)^2 2 \om} 
\coth(\frac{\bt_i \om}{2}) \tilde{S}_k
(t)   e^{i\bbox{k.r}}
\label{C10.5}\\
&=&
\frac{1}{2} \int \frac{d^2k}{(2\pi)^2 2\om} \coth(\half \bt_i \om) 
[( 1+ (\frac{\om}{W})^2) (\cosh 2Wt - 1)]
e^{i\bbox{k.r}},
\label{C10.75}
\eea
which is obtained by replacing $|\cu_k(t)|^2$ with $\tilde{S}_k(t)$
in Eqn. (\ref{B5}). 

In the limit of high initial temperatures ( $\bt \ra 0$)
and late times, $\tilde{G}(r,t)$
can be evaluated using a saddle point approximation, since the function
$ ke^{ 2W t}$
has a very sharp maximum at 
\be
k_{max} \sim \sqrt{(\frac{m_f}{2 t})}
\label{C13}
\ee
This saddle point arises from the  competition between
the phase space factor $k$ and the growth factor $W(k)t$. Note that 
this saddle point is absent in one spatial dimension.

An expansion of the integrand around this saddle point allows us to 
perform the integration at late times, leading to the 
Gaussian fall-off for the two point function,
\be
\tilde{G}(x,s) \sim 
\frac{m_f (1+L^{-2})}{4 \sqrt{2}\pi} J_0(\frac{x}{\sqrt{2 s}})
  \frac{e^{2 m_f s}}{s} e^{-\frac{x^2}{8s}},
\label{C14}
\ee
where $L \equiv \frac{m_i}{m_f}$, $x\equiv m_f r$ and $s \equiv m_f t$.

The domain sizes at late times can then be read off by looking at the ratio
\be
D (x,s) \equiv \frac{\tilde{G}(x,s)}{\tilde{G} (0,s)} \simeq 
e^{- \frac{x^2}{8 s}},
\label{C14.5}
\ee
where we have used the fact that for $s \ra \infty$, 
$J_0(\frac{x}{\sqrt{2s}} ) \ra 1$. 
Therefore, the domain size at a time $t \gg \frac{1}{m_f}$ is given by
\be
\xi (t) = \sqrt{\frac{8 t}{m_f}}.
\label{C14.75}
\ee

This scaling relation is the same as predicted by the  Cahn-Allen
equation \cite{bray} in the  classical theory of domain formation. 
The result (\ref{C14}) requires the existence of a saddle-point, which
is absent in one dimension. 
It is worthwhile noting that the Cahn-Allen equation also fails in one
dimension. 
In a sense, then, the saddle-point approximation is equivalent to the classical result.

Now as the $\Phi^4$ interaction  and the associated 
back reaction effects are ignored,
one has to be careful with this zeroth order result. In fact, the
zeroth order result has the unphysical feature of indefinite
domain growth. This shortcoming, however, cannot be overcome in a
perturbative framework, 
as emphasized by Boyanovsky et.\,al. \cite{boy} and others\cite{lanl}.
One must treat the problem non-perturbatively using, for example, 
the Hartree-Fock approximation.

To study the non-perturbative dynamics of domain formation, Boyanovsky 
et. al. \cite{boy} employed a time-dependent Hartree-Fock
approximation. 
Strictly speaking such approximations can only be justified in an 
$\frac{1}{N}$ expansion for  $O(N)$ models. This is definitely questionable 
for our case with $N=2$. Despite this shortcoming, let us proceed by 
performing the decomposition 
\be
\phi(x,t) \equiv \Phi(t) + \chi(x,t)
\label{C15}
\ee
in the Lagrangian and retaining only the terms quadratic in the ``fluctuation 
field'' $\chi(x,t)$.  Note that $\Phi(t)$ gives us the vacuum expectation 
value of the field $\phi(t)$ at time $t$, i.e.
\be
< \phi(x,t)> = \Phi(t), \qquad < \chi(x,t)> =0. 
\label{C16}
\ee 
In the Hartree-Fock approximation one assumes a condensate for 
the two-point function so that one has 
\be
<\phi^2(x,t)> \equiv \overline{\Phi^2}(t)
\label{c16.5}
\ee
The  equation of motion  for the fluctuation field is now
\be
[\frac{d^2}{dt^2} + {\bf k}^2 + m^2(t)+ \frac{\lam}{2} \overline{\Phi^2}(t) ] 
{\cu}_k(t) = 0  \quad .  
\label{C17}
\ee
We also have the condition 
\be
\Phi^2 (t) = \int d^2k |\cu_k(t)|^2 \coth ( \half \bt \om).  
\label{C18}
\ee
The set of equations (\ref{C17}) and (\ref{C18}) have to be solved 
self-consistently.

Nevertheless, one can study  the qualitative behavior 
of the solution at late times. Note that for a sudden quench,
 the equation of motion reads, for $t>0$,
\be   
[\frac{d^2}{dt^2} + {\bf k}^2 - m^2_f+ \frac{\lam}{2} \overline{\Phi^2}(t) ] \cu_k(t) 
\label{C19}
\ee
where we have chosen $\overline{\Phi^2}(0) = 0$. Since  
$\overline{\Phi^2}$ grows with time as a result of fluctuations,
the range of the unstable modes is reduced. 
At the {\em spinodal} time  $t=t_s$, defined by
\be
m^2_f = \frac{\lam}{2} \overline{\Phi^2}(t_s)
\label{C20}
\ee
there are thus no unstable modes in the system and domain growth halts.

\section{Inclusion of a finite quench}

Realistically any temperature quench occurs within a finite time. 
How does this affect the above mentioned results? To answer this
question, we will assume 
hereafter that the quench begins at time $t=0$ and ends at time 
\begin{figure}
\vspace{15pt}
\epsfxsize= 3in
\epsfysize=3in
\centerline{\epsfbox{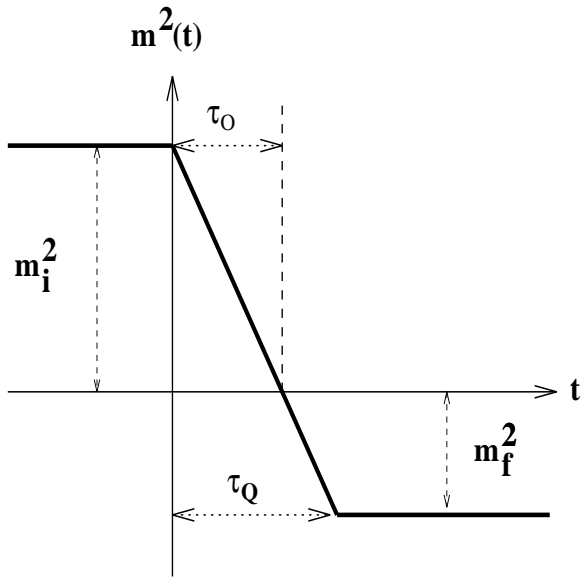}}
\vspace{20pt}
\caption{$m^2(t)$ versus   time t  for a linear quench }
\label{figure1}
\vspace{15pt}
\end{figure}
$t=\tq$ and it occurs linearly, so that the mass function is of the 
form 
\be
m^2(t) = \left\{ \begin{array}{l} m^2_i ~~{\rm for} ~~t \leq 0  \\
 m^2_i - t (\frac{m^2_i + m^2_f}{\tq}) ~~{\rm for} ~~0\leq t \leq \tq \\
 m^2_f ~~{\rm for} ~~ t \geq \tq 
\end{array} \right.
\label{D1}
\ee

Let us now define the following quantities:
\bea
x&\equiv & \frac{\tq}{m^2_i + m^2_f} \\
y&\equiv & x^{\frac{1}{3}} \omega \\
z&\equiv & x^{\frac{1}{3}} W, 
\label{D1.5}
\eea
where $\omega$ and $W$ are defined as in (\ref{C5}) and (\ref{C7.5}).
The solution to the zeroth
order equation prior to the quench,  for $ t \leq 0$, 
is given by (\ref{C5}) as before. 
During the quench period,
$ t \in [0, \tq]$, however,
the equation of motion at zeroth order reads,
\be
[\frac{d^2}{dt^2} + {\bf k}^2 + m^2_i - \frac{t}{x} ] {\cu}_k(t) = 0.   
\label{D2}
\ee
Let us now discuss the consequences of the finite-time quench.

\subsection{During the quench}

During the quench process, the solution to Eqn. (\ref{D2}) can be 
found in terms of the Airy functions, 
\be
\cu_k(t) = a_k Ai\left( x^{-\thr} (t- \om^2 x) \right) + 
b_k Bi \left( x^{-\thr} (t- \om^2 x)\right).
\label{D3}
\ee
The coefficients $a_k, b_k$ can be determined by matching 
the solutions (\ref{C5}) and (\ref{D3}) and 
their first derivatives at the beginning of the quench ,  $t=0$ :
\bea
a_k = \frac{1}{\pi} \left[ Bi~' ( -y^2 ) + i y 
Bi (-y^2 ) \right], \nonumber \\
b_k = -\frac{1}{\pi} \left[ Ai~' ( -y^2 ) + i y 
Ai (-y^2 ) \right],
\label{D5}
\eea
where the prime denotes differentiation with respect to the argument.

The domain growth function can be defined for the finite-time quench
similar to (\ref{C10}).  But, while in 
the sudden quench 
the unstable modes appear instantaneously, they will appear with   
a time lag, in the case of a finite-time quench. This can be 
understood by noting that in a linear quench the mass function
becomes negative at time ( cf. Fig. 1 )
\be
\tau_0 \equiv  \tq \left(\frac{m^2_i }{m^2_i + m^2_f} \right), 
\label{D8}
\ee
with the quench beginning at $t=0$. 
Thus the function that describes the growth of the unstable 
modes with wave-number $k$ is given by 
\be
\tilde{S}_k(t; \tau_0) \equiv |\cu_k(t)|^2 - |\cu_k(\tau_0)|^2.
\label{D9}
\ee
For an instantaneous quench there is no time lag, i.e. $\tau_0 =0$, 
and we get back the standard subtraction \cite{boy} in (\ref{C10}). 

There is another difference between a sudden quench 
and  a finite-time quench. Consider a 
 time $t$ , with $\tau_0 <t < \tq$,
and mass function  $m^2(t) = -M^2(t)$.
The phase space for the unstable modes is given by $k^2 \in [0, M^2(t)]$.
Note that this interval is smaller than that for the 
instantaneous quench, as
\be
m_f^2 - M^2(t) = (m_i^2 + m_f^2 ) [ 1- \frac{t}{\tq}] >0 \qquad 
t \in [\tau_0 ,\tq].
\label{D9.25}
\ee
 
Hence the function describing the domain growth is given by  
\be
\overline{G}_{\tau_0} (\bbox{r},t) =
\int^{k^2=M^2(t)}_0
 \frac{d^2k}{(2\pi)^2 2 \om(k)} \coth(\frac{\bt_i \om(k)}{2}) \tilde{S}_k
(t;\tau_0 )   e^{i\bbox{k.r}},
\label{D9.5} 
\ee
where the integration is performed over the unstable modes appearing
at time $t \in [\tau_0, \tau_Q]$. 

The domain size can be read off,
as before, by looking at the ratio 
\be
D_{\tau_0}(x,s) \equiv ~~\frac{\overline{G}_{\tau_0}(x,s)}{
\overline{G}_{\tau_0}(0,s)}.
\label{D12.5}
\ee 

In Figs. 2,3,4  a comparative plot of the 
domain functions for an instantaneous quench and a finite-time 
quench is given for $ \tau_0 < t < \tau_Q$.

\begin{figure}
\epsfxsize= 3.5in
\epsfysize=3.5in
\centerline{\epsfbox{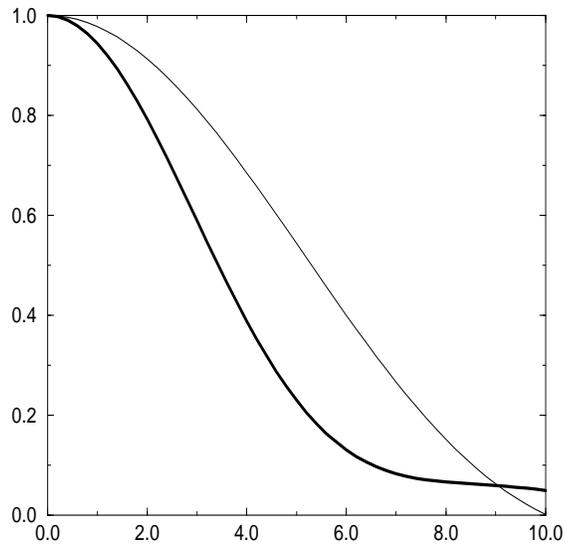}}
\caption{Comparative plot of $D(x,s)$ and $D_{\tau_0}(x,s)$ vs. $x$  
with $L \equiv .25$ at $s=.5$ with $s_Q \equiv m_f \tau_Q= 1$, $\beta
m_f=6$. The 
thin line corresponds to the case with a finite-time quench while the
thick line corresponds to the case with instantaneous quench.}
\label{figure2}
\end{figure}
\begin{figure}
\epsfxsize= 3.5in
\epsfysize=3.5in
\centerline{\epsfbox{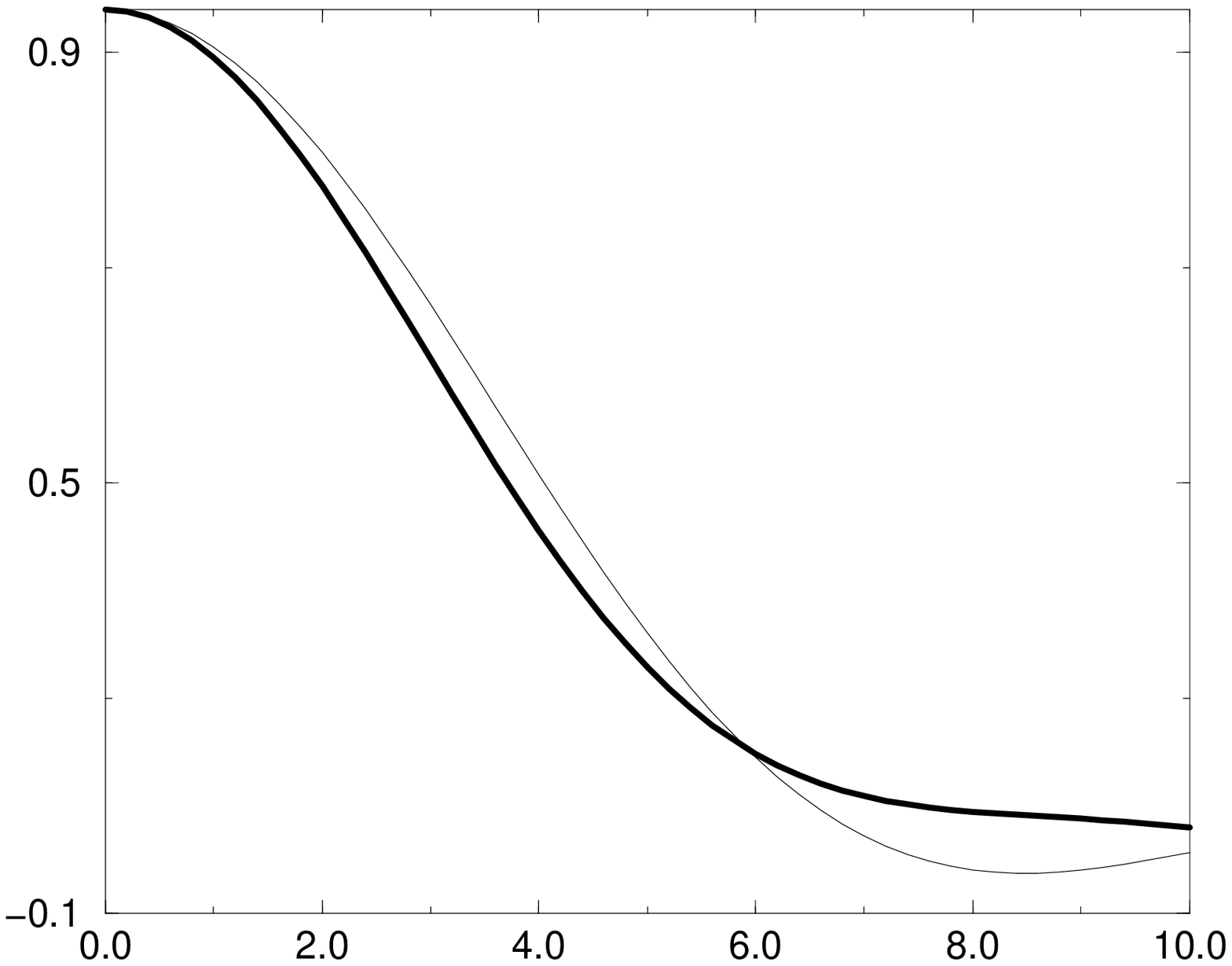}}
\caption{ Same plot as in  Fig. \ref{figure2} but with s=.7}
\label{figure3}
\end{figure}
\begin{figure}
\epsfxsize= 3.5in
\epsfysize=3.5in
\centerline{\epsfbox{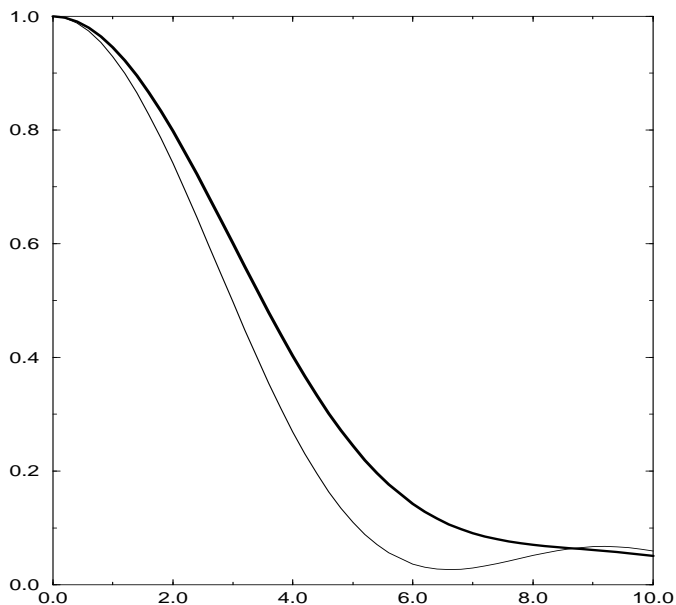}}
\caption{ Same plot as in Fig. \ref{figure2} but with s=.9}
\label{figure4}
\end{figure}

Note that during the early part of the quench, the domain sizes in the
finite-quench are larger while later the domain sizes
tend to be smaller. This can be explained as a competition between the
time lag and the reduced phase space for the unstable modes. Due to
the time lag the domains have a shorter time period to grow while due
to the smaller phase space the domain function has a larger
spread. The latter can be understood by noting that the domain
function is defined in position space as a Fourier transform.
At early times the phase space is quite small and
therefore washes out the effect of the initial time-lag. Domain 
sizes  consequently appear larger than for a  sudden quench. 
On the other hand,  at late times, the  available phase
space for the unstable modes is comparable to that for the
sudden quench. At late times the effect of  the initial
time-lag then becomes clearly visible.

\subsection{After the quench}

At the end  of the quench, the equation of motion is given by 
(\ref{C6}), with solutions of the form (\ref{C7}).

The coefficients $ A_k, B_k$ can be determined, again, by matching 
the solutions (\ref{D3}) and (\ref{C7}) and their 
first derivatives at the point $t=\tq$.
This leads to 
\bea
A_k = \half \left[ a_k \left\{ Ai(z^2) - \frac{1}{z}
Ai~'(z^2) \right\} + b_k \left\{ Bi(z^2) - \frac{1}{z}
Bi~'(z^2) \right\} \right] e^{\tq W}, \nonumber \\
B_k = \half \left[ a_k \left\{ Ai(z^2) + \frac{1}{z}
Ai~'(z^2) \right\} + b_k \left\{ Bi(z^2) + \frac{1}{z}
Bi~'(z^2) \right\} \right] e^{-\tq W}.
\label{D6}
\eea
The solution after the quench is thus given by 
\bea
\cu_k(t) = a_k &[& Ai(z^2) \cosh W(t-\tq) + \frac{1}{z} Ai~'(z^2)\sinh 
W(t-\tq)]\nonumber \\
&+& b_k [Bi(z^2) \cosh W(t-\tq) + \frac{1}{z} Bi~'(z^2)\sinh W(t-\tq)],
\label{D7}
\eea
where $a_k $ and $b_k$ are given by (\ref{D5}).
\begin{figure}
\epsfxsize= 3.5in
\epsfysize=3.5in
\centerline{\epsfbox{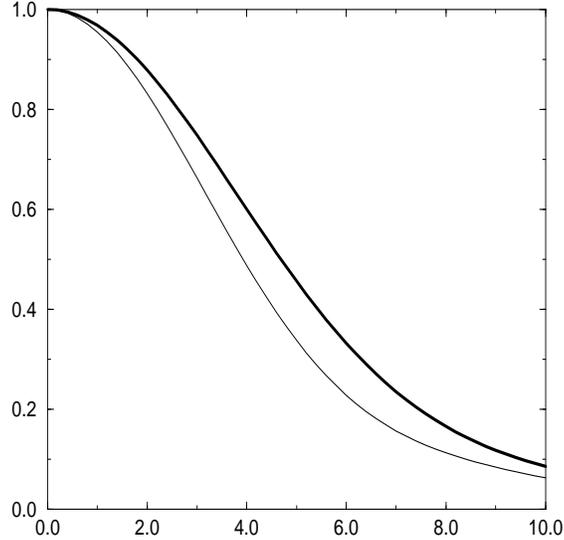}}
\caption{Same plot as in Fig.\ref{figure2} but with $s=4$}
\label{figure5}
\vspace{20pt}
\end{figure}
One can again use the domain growth function in (\ref{C10}) but with
(\ref{D7}). As the quench process is finished, however, all the
unstable modes  with $0 < k < M_f$ are present after the quench.
The domain size is given as before by (\ref{D8}). A comparative plot
is given in Fig. \ref{figure5}.  This shows the domain sizes to be smaller
compared with the sudden quench results, as a consequence of their
slower growth during the quench period.
  
This result, however,  seems
to be in contradiction with the physical picture in which a very slow
quench should not lead to appreciable domain formation. This puzzle
points out that we have to take into account the effects of the back-reaction.

\subsection{ The role of Back-Reaction}

Let us now incorporate the back-reaction into the picture with a
finite-time quench.  The equation 
of motion in the Hartree-Fock theory, during the quench period,  would
read as
 
\be
[\frac{d^2}{dt^2} + {\bf k}^2 + m^2_i+ \frac{\lam}{2}
\overline{\Phi^2}(t) - \frac{t}{x} 
] \cu_k(t). 
\label{D13}
\ee
Solving this equation is difficult. To
gain some qualitative understanding, let us again consider an adiabatic
quench. Note that as $\Phi (t=0) =0$ there is not much change in the initial 
behavior of system compared to the zeroth order case. 
As $\Phi(t)$ grows, however, 
one has  $\Phi^2(t) > 0$ and the ``back-reaction'' competes with the
decreasing  mass-function, producing a slower quench. 
Recall that domain formation is associated with
the appearance of unstable modes and due to the slower quench they
will appear even later compared to the zeroth order case. This is
definitely true for  small values of $\lam$, so that   
the mass function {\it can} pass through zero at  late times. By our
earlier arguments, one can see that the 
domains formed subsequently will be larger again.  On the other hand,
for larger values of $\lambda$, if $\Phi$ grows fast, the effective
mass-function might not become negative at all 
during a weak quench and there will be no domain formation. 
Thus, if defects are to form  
one would need a quench rate faster than the growth rate for the field
$\Phi$, as envisaged in the picture of Zurek \cite{zur}.

\section{ The quench on the Goldstone Modes }

The Halperin-Mazenko-Liu defect density formula 
involves the complete two-point function whereas we have been dealing with a
restricted two-point function which only involves the 
long-wavelength unstable modes. On the other hand, defects are 
localized objects and
thus they carry short-wavelength modes with them. In fact, the O(2) global
defect is associated with a phase variable which has no apparent
cutoff associated with it, as we shall see below. 

In our above treatment of the $O(2)$ symmetric scalar model, we have used a
Cartesian parametrization for the fields. 
The zeroth order equations of motion for the two fields accordingly decouple and appear as linear
equations. The phase degree of freedom (the Goldstone mode of the
theory) nevertheless interacts with the radial mode. The defect 
configurations in the $O(2)$ model are associated with windings of
this Goldstone phase.
It can therefore be  misleading to use the Cartesian  parametrization
when studying defect formation. Let us instead rewrite the O(2) model 
in terms of a complex field $\Phi$ (polar decomposition): 
\be
\Phi \equiv \phi_1 + i \phi_2 \equiv F e^{i \Theta }.
\label{E1}
\ee
In terms of $F$ and $\Theta$, the Lagrangian now reads 
\be
L = \frac{1}{2} \left[ \partial_\mu F \partial^\mu F + F^2
\partial_\mu \Theta \partial^\mu \Theta \right] - \frac{1}{2} m^2(t)
F^2  - \frac{1}{4} \lambda F^4, 
\label{E2}
\ee
exhibiting the {\em Goldstone} nature of the phase $\Theta$ and its
coupling to the radial mode $F$.

The equations of motion following
from the Lagrangian (\ref{E2}) are given by:
\bea
\partial_\mu \partial^\mu  F + \left[ m^2 (t) + \lambda F^2 - ( \partial_\mu
\Theta \partial^\mu \Theta)  \right] F=0 \label{E2.5} \\
\partial_\mu ( F^2 \partial^\mu \Theta) =0
\label{E3}
\eea
Note that in (\ref{E2.5}) the term involving the Goldstone
modes appears with the opposite sign to the time-dependent mass function
$m^2(t)$, in contrast to the case with back-reaction. 
Note also that the quench does not affect the Goldstone modes 
{\it directly}. 

We will attempt to follow the dynamics of this system by using a
Born-Oppenheimer like approximation, treating $\Theta$ as a {\em fast}
variable and $F$ as a {\em slow} variable.  

If we ignore the back-reaction due to the field $\Theta$,
 the VEV for the radial field $F$ is given  at
any instant during the quench by
\be
\lambda F^2 (t) =\left\{ \begin{array}{l} 0 \quad (t=0) \\
 - m^2(t) + ( \partial_\mu \Theta \partial^\mu \Theta) \quad (t > 0)
\end{array} \right.
\label{E4}
\ee
   
Using (\ref{E4}) in the second equation of (\ref{E3}) one gets {---}
\bea
&&\lambda \partial_\mu ( F^2 \partial^\mu \Theta) \nonumber \\
&&= \partial_\mu \left[ (m^2(t) + ( \partial_\nu
\Theta  \partial^\nu \Theta) ) \partial^\mu \right] \Theta =0
\label{E5}
\eea
One can attempt to simplify the dynamics by ignoring the
 $(\partial_\nu \Theta \partial^\nu \Theta)$ term in the above equation
and solve instead the equation,
\be
\partial_\mu \left( m^2(t) \partial^\mu \Theta \right) = 0,
\label{E6}
\ee
where we have dropped the extra term from the equation of motion 
as it is less relevant at long-wavelength. 
Equation (\ref{E6}) can be expanded as 
\be
\ddot{\theta} + \frac{1}{m^2}(\frac{d}{dt}m^2 )\dot{\theta} - 
\nabla^2 \theta = 0.
\label{E7}
\ee

The second term can be interpreted as a ``friction'' term which changes sign
from positive to negative as the mass function changes
sign. Accordingly, there will be production of Goldstone particles
which will feed into the dynamics of the $F$ field, as in
(\ref{E2.5}). This is similar to the picture in \cite{devega}.    
Since there is  no intrinsic mass scale for $\theta$
field one needs to consider all frequencies including the high
ones. This is also necessary as the defects are localized objects
which can only be probed with high frequencies.  

Note that when $m^2 (t)  \ra 0$ , only the second term which involves  
the first derivative in time dominates,
\be
2 \dot{m} \dot {\theta} \approx 0 \qquad \mbox{ for $m \ra 0$}. 
\label{E8}
\ee
In the linear quench this shows that the Goldstone modes become 
time-independent immediately prior to the appearance of the unstable
modes. Thus the defects get
disentangled from the evolution of the field $F$, as in the
picture proposed by Zurek \cite{zurek}.
   
An adiabatic solution to Eqn. (\ref{E7}) is given by
\bea
&&\theta ( x, t ) = \int d^2k d \omega [ \theta ( \omega,k ) +
\mbox{c.c.} ] , \nonumber \\
&& \theta (\omega, k ) = e^{i {\bf k}\cdot {\bf x} } e^{-i
\omega t}  
\label{E8.5}
\eea
where 
\be
\omega \approx \pm ( k^2 - \frac{1}{4 m^4 x^2})^{\frac{1}{2}} -
\frac{i}{2m^2x}
\label{E8.75}
\ee
in the adiabatic approximation, with $x = \frac{\tau_Q}{m_i^2 +
m_f^2}$. For the static situation, one has  $x=\infty$ and one gets
back the standard free-particle. On the other hand, when $m^2 \ra 0$,
we find $\omega \ra 0, -i \infty$ leading to the static situation
discussed above. 

Though qualitatively the above scenario 
supports the picture of defect formation, there is
definitely much work  needed to get  quantitative
predictions. Specifically, the
domain sizes can be found from the two-point function for the $F$
field. Because of the coupling to the Goldstone modes the evaluation
of this two-point function is quite involved.
One alternative would be to integrate
out the Goldstone field $\Theta$ ( after a Gaussian integration) 
leading to an effective action involving the $F$ field only

\bea
L_{eff}(F) &=& L_0 -\frac{1}{2} \tr \ln (\partial_\mu F^2
\partial^\mu) \nonumber \\
&=& L_0  -\frac{1}{2} \tr \ln (F^2 \partial^2 ) - \frac{1}{2} \tr \ln
( 1+  \frac{\partial_\mu F^2 \partial^\mu}{F^2 \partial^2}) + \cdots.
\label{co1}
\eea
Note that in the second line we have expanded the logarithm. Note  
that the operators $F^2 \partial^2 $ and $ \partial_\mu F^2 \partial^\mu$
do not commute and therefore in the above  expansion 
(using the Baker{--}Campbell{--}Hausdorff formula) there are other terms which are 
represented by the ellipsis. The second term
basically cancels the dependence on $F$ in the functional measure in
the path integral. Though naively it seems that  the third term in
(\ref{co1}) suffers from an infrared divergence, the  finite temperature 
of the heat bath provides us with an infrared cutoff. In light of this
one can attempt to expand the logarithm in the third term to get an effective
action for $F$ which takes into account of the ``backreaction'' of the
Goldstone modes. In practice this is rather complicated as the
the functional determinant is time-dependent (due to the
time-dependence of the cutoff). For an initial high temperature, the
logarithm can be expanded as $\frac{\partial_\mu F
\partial^\mu}{F^2\partial^2} \sim \frac{\partial_\mu F
\partial^\mu}{F^2 T_i^2}$ is small. During the later stages of the
quench, however, this may no longer be true. 
It might be more appropriate, therefore,  to use the recently
developed formalism of the non-equilibrium effective action
\cite{wetterich} since it generates time-dependent correlation
functions.

\section{Conclusions and Remarks }

In this letter, we have examined the formation of defects in phase
transitions induced by a finite-time quench on $O(2)$ symmetric scalar
field theories.  We find that adopting the treatment of Boyanovsky
et. al.  naively with a  Cartesian  parametrization leads, to the
zeroth order, to smaller
defect densities during the early quench period while higher defect
densities for the later phase. This shows that the problem of the
back-reaction needs to be studied properly. 

We also propose that a more physical picture will be found by adopting
the polar parametrization of the group manifold.
This makes explicit the interaction of the true Goldstone phase field with the
time-dependent radial field. In the polar parametrization the theory becomes interacting
even in the absence of the $\Phi^4$ interaction.  This makes the
analysis of the model much harder. 
For a finite time quench, however, such a parametrization suggests qualitative
support for the Zurek picture of defect formation. This
will be valid for the $O(n)$ symmetric scalar models as well. 
One can attempt to solve the time-dependent model numerically. 
We hope to report on this in the near future.

\section*{Acknowledgment}

We would like to thank I.J.R. Aitchison, M. Falcioni, A. Kovner, 
M.C. Marchetti and R.J. Rivers for discussions. A.M. also 
expresses gratitude to M. Harada for his generous help with 
MATHEMATICA. This work was
supported by the US Department of Energy under contract number 
DE-FG02-85ER40231 and PPARC (U.K.).

 \appendix

\section*{Real-time finite temperature field theory}

The real time formalism ( also known as the Schwinger-Keldysh 
formalism) allows one to compute {\it in-in} matrix elements,
and hence is useful for computing time-dependent averages. 
For a review of finite temperature field theory in this 
formalism see, for example,  \cite{LeBelac}. Here  we only quote the
formulae needed for our purposes.

In this paper, we will 
be interested in the equal-time two-point Green  function   
\be
G(x,t) \equiv ~< \phi(x,t) \phi(0,t)> ~= \tr [ \rho_0 \phi(x,t) \phi(0,t)].
\label{B1}
\ee
Here $\rho_0$ is the thermal density matrix describing the 
initial condition,
\be
\rho_0 = \frac{ e^{-\bt H_i}}{ \tr  e^{-\bt H_i}}\,,
\label{B2}
\ee
with  $H_i$ the Hamiltonian describing the 
system initially. 

For a bosonic field in $D$ spatial dimensions with a mass $m$ 
we have the standard  expansion:
\be
\phi(x,t) = \int \frac{d^Dk}{(2\pi)^D 2 \om(k) } [ a_k \cu_k(t) + 
a^\dagger _k {\cu_k(t)}^\dagger ] e^{i \bbox{k.x}}
\label{B3}
\ee
where $\om(k) =( k^2 + m^2)^\half $ and  $\cu_k(t)$ is the one-particle 
wavefunction in momentum space.
The particle creation and destruction operators
satisfy 
\be
[ a_k , a^\dagger_{k'}] = (2\pi)^D 2 \om (k) \delta^D ( k - k').
\label{B4}
\ee
Substituting (\ref{B3}) in (\ref{B1}) one gets,
\be
G(x,t) = \int \frac{d^Dk}{(2\pi)^D 2 \om(k) } \coth ( \frac{\bt \om(k)}{2})
|\cu_k(t)|^2.
\label{B5}
\ee

\end{document}